\documentclass[preprint,a4,aps]{revtex4}

\begin{document}
\title{Noncommutative Relativistic $U(N)$ Chern-Simons Solitons}


\author{Ee Chang-Young${}^{1,}$\footnote{cylee@sejong.ac.kr}}
\author{Kyoungtae Kimm${}^{1,}$\footnote{helloktk@naver.com}}
\author{Daeho Lee${}^{1,}$\footnote{dhlee@sju.ac.kr}}
\author{Youngone Lee${}^{2,}$\footnote{youngone@daejin.ac.kr}}

\vspace{10mm}

\address{${}^{1}$Department of Physics and Institute of Fundamental Physics\\
Sejong University, Seoul 143-747, Korea\\
${}^{2}$Department of Physics, Daejin University,
Pocheon, Gyeonggi 487-711, Korea}


\begin{abstract}
We investigate BPS soliton solutions of $U(N)$ Chern-Simons gauge theory coupled
to a scalar field in  noncommutative plane.
With a scalar field in the fundamental representation, we show that the BPS equation becomes that of abelian Chern-Simons theory in the unitary gauge.
We also find a class of particular solutions for the BPS equation with scalar field in the adjoint representation.
\end{abstract}
\maketitle
The study of solitonic solutions on noncommutative spaces has
 been  an active topic over the past years because of their possible relevance to strings and brane dynamics~\cite{harvey_102076,douglas_2001,szabo_2003}.
So far, the noncommutative counterparts of variety of non-perturbative solutions like instantons, monopoles, vortices in  commutative field theories have been found~\cite{nekrasov_1998,hashimoto_1999,jatkar_2000}.
Extension to noncommutative space supplies further solitonic configurations
whose commutative counterparts do not exist.

In 2+1 dimensions, the Chern-Simons term  plays an important role in  condensed matter physics. Chern-Simons theory on   noncommutative plane can describe the quantum Hall system~\cite{susskind_1029}.
The soliton solutions of relativistic and nonrelativistic Chern-Simons theories coupled with matters have been studied extensively.
Various noncommutative extensions of these solutions
were also explored by many authors~\cite{bak_25018,lozano_2003}.
In~\cite{klee_2004}, general construction of nonrelativistic BPS solitons of nonabelian Chern-Simons theory coupled with adjoint matter was conjectured.
This conjecture is based on the fact that the BPS equations for Chern-Simons solitons on noncommutative plane
can be related to the equations of the $U(N)$ noncommutative chiral model which
 can  also be solved formally by the Uhlenbeck¡¯'s uniton method~\cite{uhlenbeck}.

In the present paper we investigate  noncommutative BPS solitons in $U(N)$ Chern-Simons theory coupled to a scalar field.
We express the scalar field both in the fundamental and adjoint representations.
In~\cite{klee_1991},
the BPS solitons for $SU(N)$ gauge group in commutative plane were studied in detail.

The noncommutative plane with spatial coordinates $(x,y)$ is defined by the following commutation relation:
\begin{equation}
[x, y] = i \theta, ~~\theta>0.
\label{comm_rel}
\end{equation}
Field theory on this noncommutative plane can be realized
by replacing the ordinary pointwise products in the corresponding commutative field theory with the Moyal $*$-product.
The Moyal $*$-product is defined by
\begin{eqnarray}
(f*g)(x) &=& e^{ \frac{i}{2}\theta^{\mu\nu}
\frac{\partial}{\partial x^\mu} \frac{\partial}{\partial y^\nu}}f(x)g(y)
|_{x=y}  \cr
&=&f(x)g(x)+ \frac{i}{2} \theta^{\mu\nu}
\frac{\partial f}{\partial x^\mu}
\frac{\partial g}{\partial x^\nu}+ {\cal O}(\theta^2),
\end{eqnarray}
with commutative coordinates $x$ and $y$.

Noncommutativity with the commutation relation (\ref{comm_rel}) can be equivalently expressed with operators on a certain Fock space as follows.
We define $a$ and $a^\dagger$ as,
\begin{equation}
a :=\frac{x+iy}{\sqrt{2\theta}},   
~~a^{\dagger} :=\frac{x-iy}{\sqrt{2\theta}},
\end{equation}
then $[a,a^\dagger]=1$.
This  satisfies  the algebra of quantum harmonic oscillator.
Here we use this Fock space for the representation of the algebra~(\ref{comm_rel}).
Now, the spatial integration becomes the trace over the Fock space,
$\int d^2x \rightarrow 2\pi \theta {\rm Tr}$,
and the spatial derivatives become
\begin{equation}
\partial_{-}:=
\partial_1 -i \partial_2
\rightarrow -\sqrt{\frac{2}{\theta}}[ a^\dagger, ~],
~~\partial_{+}:=
\partial_1 + i \partial_2
\rightarrow \sqrt{\frac{2}{\theta}}[ a, ~].
\end{equation}

\section*{Fundamental Representation}
First, we consider a noncommutative extension of  Chern-Simons theories coupled to a scalar field in the fundamental representation.
With $U(N)$ gauge group we study solitonic solutions in the relativistic case.
The gauge field dynamics is described by the Chern-Simons Lagrangian defined as
\begin{equation}
L_{CS} = \kappa
\epsilon^{\mu\nu\rho}
{\rm{tr}}\left(
A_\mu \partial_{\nu}A_{\rho} - \frac{2}{3} i A_\mu  A_\nu  A_\rho
\right),
\end{equation}
where $A_\mu$ are given by $N \times N$ hermitian matrices
and ``tr" means trace over the gauge group.

We consider a  model coupled to a scalar field  defined by the following Lagrangian
\begin{equation}
L = L_{CS}
+| D_\mu\phi|^2 - V(\phi, \phi^\dagger),
\label{fund_lag}
\end{equation}
where $D_\mu \phi = (\partial_\mu - iA_\mu ) \phi$.
The potential for the fundamental scalar field is chosen to be
\begin{equation}
V(\phi, \phi^\dagger) = \frac{1}{4\kappa^2}|\phi|^2 (|\phi|^2- v^2)^2,
\label{fund_pot}
\end{equation}
where $|\phi|^2 = \phi^\dagger \phi$.
This potential allows  soliton solutions which  saturate the BPS bounds.
Note that unlike our case the commutative $SU(N)$ Chern-Simons model with a fundamental scalar studied in~\cite{klee_1991} has the same form of potential which has explicit dependence on flavour $N$.
The potential has two degenerate vacua: the symmetric phase where $\phi=0$ and the asymmetric phase where $\phi=v$.

Using the time translational invariance of the model, the Hamiltonian can be constructed as
\begin{equation}
H = 2\pi\theta{\rm Tr} \left(
|D_0 \phi|^2 + |D_i\phi|^2 + V(\phi , \phi^\dagger )
\right).
\end{equation}
This model is invariant under the global $U(1)$ phase shift of scalar field in addition to $U(N)$ gauge symmetry.
The corresponding conserved global charge is given by
\begin{equation}
Q = i 2\pi\theta{\rm Tr}\left[(D_0 \phi)^\dagger \phi - \phi^\dagger (D_0 \phi )\right].
\label{global_charge}
\end{equation}
The Gauss law derived from (\ref{fund_lag}) is
\begin{equation}\kappa F_{12}^a
=i\sum_{\alpha,\beta=1}^{N} T^a_{\alpha\beta}
\left( \phi_\beta (D_0 \phi)^\dagger_\alpha
+ (D_0 \phi)_\beta \phi^\dagger_\alpha \
\right),
\label{fund_gauss}
\end{equation}
where $T^a, a=0,1,2,...,N^2-1$ are $U(N)$ gauge group generators. We use explicit matrix representation of the generators as follows:
$T^0=\frac{1}{\sqrt{2N}}\textrm{diag}(1,1,...,1)$ which is the global $U(1)$ generator,
$T^D := T^{N^2-1}= \frac{1}{\sqrt{2N(N-1)}}\textrm{diag}(1,...,1,N-1)$, and for the remaining generators we follow the convention in Ref.~\cite{klee_2004}.

We are interested in finding a static configuration with the lowest energy
for a given global charge $Q$.
With (\ref{global_charge}), the energy functional can be written as complete squared forms plus a global charge:
\begin{equation}
H = 2\pi\theta{\rm Tr} \left(
\left| D_0 \phi \pm \frac{i}{2\kappa} \phi ( | \phi |^2- v^2 )\right|^2
+ |(D_1 \pm i D_2) \phi | ^2
\right)
\pm \frac{v^2}{2\kappa} Q.
\label{energy_func}
\end{equation}
The energy is bounded from below by $\frac{v^2}{2\kappa}|Q|$, and the field configuration saturating the  bound should satisfy
the following BPS equations,
\begin{eqnarray}
(D_1 \pm i D_2 )\phi & =& 0,\cr
D_0 \phi \pm \frac{i}{2\kappa} \phi ( |\phi|^2 - v^2 ) &=&0,
\label{fund_sd}
\end{eqnarray}
as well as the Gauss constraint~(\ref{fund_gauss}).
The above equations imply that
the minimum energy configuration is stationary in time.

To solve the BPS equations, let us consider a semi-unitary gauge
\begin{equation}
\phi^T  = (0,..., 0, \chi),
\label{semi-unitary}
\end{equation}
where $\chi$ is a complex scalar field.
In this gauge, the  nontrivial components of gauge field are $A_\mu^0$
and $A_\mu^D$ only. Furthermore, one can notice from~(\ref{fund_sd})
that
\begin{equation}
A_\mu^D = - \sqrt{N-1} A_\mu^0.
\end{equation}
We define a new gauge field $V_\mu$ as
\begin{equation}
V_\mu := \sqrt{\frac{2}{N}} A_\mu^0.
\label{new_gauge}
\end{equation}
With the Gauss law constraint (\ref{fund_gauss}) and $V_{\pm}:= V_1 \pm i V_2$, the BPS equations in (\ref{fund_sd}) become
\begin{eqnarray}
(\partial_\pm -i V_\pm )\chi &=& 0,\cr
B \pm  \frac{1}{2\kappa^2} \chi (\chi^\dagger \chi -v^2 )\chi^\dagger &= &0,
\label{red_sd}
\end{eqnarray}
where $ B:=\partial_1 V_2 -\partial_2 V_1 -i[V_1, V_2]$.
In the commutative $SU(N)$ model considered in Ref.~\cite{klee_1991},
full BPS equations reduced to those of the $U(1)$ model.
A solution of BPS equations in the unitary gauge was obtained by embedding the $U(1)$ vortex solution.
Since the same reduction occurs here, we  can use   the same method.

Now, we investigate   rotationally symmetric configurations of BPS equations explicitly. {}From the gauge transformation property we can decompose the gauge field into gauge covariant part $K$ and non-covariant one as in~\cite{bak_125010}:
\begin{equation}
V_{-}  = i \sqrt{\frac{2}{\theta}} (a^\dagger - K^\dagger ).
\label{cov_gauge}
\end{equation}

We consider the BPS equation (upper sign in~(\ref{red_sd}))
and take the following
ansatz for $\chi$ and $V_\mu$ (via $K$)
\begin{equation}
\chi = v\sum_{n=0}^{\infty} \chi_n | n\rangle \langle {n+m}|,
~~
K = \sum_{n=0}^{\infty} k_n |n\rangle \langle n+1| ,
\label{fund_sd_sol1}
\end{equation}
for a given positive integer $m$.
Inserting these into the BPS equations we get the following algebraic recurrence equations for coefficients $\chi_n$ and $k_n$:
\begin{eqnarray}
k_n  &=& \sqrt{n+m+1} \frac{ \chi_n}{ \chi_{n+1}}, ~~n=0,1,2,...,\cr
\chi_1^2 &=& \frac{\chi_0^2 (m + 1) }{1+ \eta \chi_0^2(1- \chi_0^2) },\cr
\chi_{n+1}^2 &=&
\frac{(n+m+1)\chi_n^2}{1+\eta \chi_n^2(1-\chi_n^2) + (n+m)\chi_{n-1}^2 /\chi_n^2}, ~~n=1,2,...
\label{fund_rec1}
\end{eqnarray}
where $\eta = \theta v^4/2\kappa^2$. These equations enable us to find all $k_n$'s and $\chi_n^2$'s for given $k_0$ and $\chi_0^2$.
The energy of the configuration is given by the global charge $Q$ which can be expressed by the magnetic flux of the gauge field $V_{\mu}$:
\begin{equation}
Q = 4\pi \theta \kappa {\rm Tr} (B).
\end{equation}
With the above ansatz, the magnetic field $B$ is given by
\begin{equation}
B = \frac{1}{\theta}\sum_{n=0}^{\infty} ( k_n^2 - k_{n-1}^2-1 ),
\end{equation}
where $k_{-1}=0$.
Hence, with Eq.~(\ref{fund_rec1}) one can easily notice that in order to have a
finite energy configuration, the asymptotic values of $k_n$ and $\chi_n$ should
meet the following conditions:
\begin{equation}
k_n^2\rightarrow (n+m+1), ~~~
\chi_n\rightarrow 1 ~~{\rm as}~~n\rightarrow \infty.
\end{equation}
Since the number operator $N:=a^\dagger a$ is related to the radial distance $r^2 = x^2+y^2 = (\theta/2)N$,
the above conditions mean that the scalar field should remain
at asymmetric phase
and the gauge field becomes pure gauge at spatial infinity.
The energy of BPS configuration becomes
\begin{equation}
H = 2\pi v^2 m.
\end{equation}

%

In the commutative case, parity transform of BPS solution gives
the corresponding anti BPS solution.
However, due to  parity violation in   noncommutative field theory, we cannot use the procedure of the commutative case.
The rotationally symmetric configuration ansatz for the anti-BPS equations (lower sign) reads
\begin{equation}
\chi = v \sum_{n=0}^{\infty}
\chi_{n} |n+m\rangle\langle{n}| ,
~~K = \sum_{n=0}^\infty k_n |n\rangle\langle{n+1}|,
\label{fund_sd_sol2}
\end{equation}
for a given positive integer $m$.
Again with these ansatz one can find the following recurrence relations without difficulty.
\begin{eqnarray}
k_{n} &=& \sqrt{n+1}, ~~n=0,1,...,m-1,\cr
k_{n} &=& \sqrt{n+1-m} \frac{\chi_{n+1}}{\chi_{n}},~~n=m,m+1,...\cr
\chi_1^2 &=& \chi_0^2 (1+m +\eta \chi_0^2(\chi_0^2-1)),\cr
\chi_{n+1}^{2} &=&\chi_n^2 (1+n\chi_n^2/\chi_{n-1}^2 + \eta \chi_n^2(\chi_n^2-1))/(n+1),~~n=1,2,,
\label{anti_sd_rec}
\end{eqnarray}
The energy of this anti-BPS configuration is
\begin{equation}
H = 2\pi v^2 m.
\end{equation}
It is possible to show that for a small $\theta$ value, the solutions (\ref{fund_sd_sol1}) and (\ref{fund_sd_sol2}) approach to
the well-known solutions of commutative (anti-)BPS equations~\cite{lozano_2003}.
Noncommutativity enables us to find an unusual solution which has no
commutative counterpart.
From the third equation of (\ref{anti_sd_rec}) we see that
if $\chi_{0}^2 =(v^2/2)\left(1 \pm   \sqrt{1- 4 (m+1)/\eta} \right)$,
then $\chi$ consists of a single term $\chi\sim|m\rangle\langle{0}|$.
However, with truncated expression $\sum_{n=0}^{m-1}\sqrt{n+1}|n\rangle \langle{n+1}|$ for $K$ the Gauss's law cannot be satisfied.
Fortunately, in Ref.~\cite{bak_25018} an exact solution  was found
as
\begin{equation}
\chi =\chi_\pm|m-1\rangle \langle{0}|,
~~K = P_{m} c P_{m} + S_m c S_m^\dagger, ~~m=1,2,...
\end{equation}
where $P_m = \sum_{n=0}^{m-1} |n\rangle\langle{n}|$ is a projection operator and
$S_m =\sum_{n=0}^\infty |n+m\rangle\langle{n}|$ is a shift-operator. The constant $\chi_\pm$ is given by
\begin{equation}
\chi_\pm^2  = \frac{v^2}{2}\left( 1 \pm \sqrt{1- 8\kappa^2 m/\theta v^4}\right).
\end{equation}
Since the solution exists only for  $\theta \ge 8\kappa^2 m/ v^4$,
this solution has no commutative counterpart.
Note that the magnetic field of $V_\mu$ is given by
\begin{equation}
B = - \frac{1}{\theta} m |m-1\rangle\langle{m-1}|,
\end{equation}
and the energy of the configuration is given by
\begin{equation}
H =  2\pi v^2 m.
\end{equation}

\section*{Adjoint Representation}
Next, we consider the scalar field in the adjoint representation of
the gauge group $U(N)$.
The dynamics of the gauge  and scalar fields is governed by the following Lagrangian:
\begin{equation}
L  = L_{CS} + {\rm tr} |D_\mu \phi |^2   - V(\phi, \phi^\dagger),
\label{adjoint_lag}
\end{equation}
where $D_\mu\phi = \partial_\mu \phi - i[A_\mu, \phi ]$.
If we choose a sextet order potential
\begin{equation}
V(\phi, \phi^\dagger) = \frac{1}{4\kappa^2} {\rm tr}
\left|
\left[ [ \phi, \phi^\dagger], \phi\right] -v^2 \phi
\right|^2,
\end{equation}
the model also enjoys the BPS equations.
The energy functional is given by
\begin{equation}
H = 2\pi\theta \textrm{Tr}
\left(  \textrm{tr}
\left|D_0 \phi
\pm \frac{i}{2\kappa}\left(\left[[ \phi,\phi^\dagger],\phi\right]-v^2\phi\right)
\right|^2
+\textrm{tr}\left|(D_1 \pm i D_2 )\phi \right|^2
\right)
\pm \frac{v^2}{2\kappa} Q.
\end{equation}
The conserved global $U(1)$ charge for the adjoint scalar case is given by
\begin{equation}
Q = -i 2\pi \theta \textrm{Tr} \left[ \textrm{tr}
\left( (D_0\phi)^\dagger \phi
-\phi^\dagger (D_0\phi) \right)
\right].
\end{equation}
The Gauss law constraint for this model is
\begin{equation}
F_{12} =-\frac{i}{2\kappa} \left(
[D_0 \phi, \phi^\dagger ] - [\phi, (D_0\phi)^\dagger ] \right).
\end{equation}
The saturation of energy functional occurs when the following BPS equations
are satisfied
\begin{eqnarray}
(D_1 \pm i D_2 )\phi & =& 0, \cr
D_0\phi\pm \frac{i}{2\kappa}\left(\left[[ \phi,\phi^\dagger],\phi\right]-v^2\phi\right) &=& 0.
\end{eqnarray}
With the Gauss law constraint, the BPS equations become
\begin{eqnarray}
D_\pm \phi &=& 0,\cr
F_{+-}  &=& \mp \frac{i}{\kappa^2} \left[v^2 \phi -[[\phi, \phi^\dagger],\phi],\phi^\dagger \right].
\label{adjoint_sdeq}
\end{eqnarray}

Before we solve the above BPS equations, we review the solutions of
the nonrelativistic DJPT model on noncommutative plane~\cite{klee_2004}
where the BPS equations read as:
\begin{eqnarray}
D_\pm \phi &=& 0 ,\cr
F_{+-} &=&  i [\phi, \phi^\dagger ].
\label{nonrel_sdeq}
\end{eqnarray}
In the commutative case, there is an equivalence between the BPS equations and the chiral model equations.
The solutions of chiral model have been completely classified in~\cite{leznov1983}.
In the noncommutative case, it was conjectured  in~\cite{klee_2004} that
the general solution of chiral model with finite energy can be constructed with
any holomorphic  projection operators.

Due to the appearance of the triple commutator in (\ref{adjoint_sdeq})
we cannot directly apply the method suggested in~\cite{klee_2004}.
However note that
the vacuum configuration is given by
$[[\phi, \phi^\dagger],\phi] = v^2 \phi$,
and if
the field configuration satisfies
\begin{equation}
[[\phi, \phi^\dagger],\phi] \propto \phi,
\label{phi_constraint1}
\end{equation}
then we end up with the same type of BPS equations appeared in~\cite{klee_2004}.
I.e., the BPS equations become those of the nonrelativistic DJPT model.
Thus the solutions found in~\cite{klee_2004} would become solutions in our case if
they satisfy the additional
condition~(\ref{phi_constraint1}).
Only a part of the solutions for~(\ref{nonrel_sdeq}) satisfies (\ref{phi_constraint1})
as we see below.
One can check that the following form of holomorphic projection operator
${\cal P}$
satisfies the condition~(\ref{phi_constraint1}):
\begin{equation}
{\cal P}=\left(
\begin{array}{cccc}
d_1 &        &       		&  \\
    & \ddots &       		&  \\
    &        &  d_{N-1}     &  \\
    &        &       		& \sum_{n=0}^{m-1}|n\rangle\langle{n}| \\
\end{array}
\right), ~~d_1=...=d_{N-1}=0 \textrm{ or }1,
\label{proj_op}
\end{equation}
where $m=1,2,...$.
From the relation $\phi = -\sqrt{\frac{2}{\theta}}[a^\dagger,{\cal P}]$
in the DJPT model for BPS soliton (upper sign),
$\phi$ takes the form:
\begin{equation}
\phi =\textrm{diag}(0,...,0,h),
\label{ad_sol1}
\end{equation}
where
\begin{equation}
h = h_0 |m\rangle\langle{m-1}|.
\label{ad_ansatz}
\end{equation}
The constant $h_{0}$ is introduced to satisfy (\ref{phi_constraint1})
and   determined from the BPS equations
(anti-BPS solution can also be obtained with the relation $\phi = \sqrt{\frac{2}{\theta}}[a,{\cal P}]$).
Now one important comment is in order. In fact, one can transform the projection operator in (\ref{proj_op})
into that of Ref. \cite{klee_2004} by unitary transformation with an appropriate unitary operator $U$ \cite{hadasz_2001}.
Namely, the  $\phi$ given by  $\phi = -\sqrt{\frac{2}{\theta}}[a^\dagger , U{\cal P}U^\dagger]$
satisfies (\ref{nonrel_sdeq}).
However, this does not satisfy the extra constraint (\ref{phi_constraint1}), thus
 cannot be our solution.
On the other hand, the $\phi$ given by  $\phi = -\sqrt{\frac{2}{\theta}}[U a^\dagger U^\dagger, U{\cal P}U^\dagger]$
satisfies the nonrelativistic BPS equation (\ref{nonrel_sdeq}) as well as the extra constraint as we saw above.
This conforms with our previous comment that only a part of the solutions of \cite{klee_2004} becomes our solutions.
{}From~(\ref{adjoint_sdeq}) and~({\ref{ad_sol1}), one can   notice that
the nonvanishing components of $U(N)$ gauge field are $A_\mu^0 $ and $A_\mu^D$.
The nature of this solution finding process is similar to the abelian embedding
in the previous case with fundamental scalar.
Inserting the $\phi$ in~(\ref{ad_sol1}) into the BPS equations~(\ref{adjoint_sdeq}) yields
\begin{equation}
h = -\frac{v}{2}
\left( 1\pm \sqrt{ 1- \frac{16\kappa^2 m}{\theta v^4}}\right)^{1/2}
|m\rangle\langle{m-1}| .
\end{equation}
This kind of solutions does not exist in the commutative limit ($\theta\rightarrow 0$)
since the parameter $\theta$  should satisfy $\theta > 16\kappa^2 m/v^4$.
When $\theta$ is finite, $h$ does not go to its vacuum.
This is different from the commutative case.
The  conserved $U(1)$ charge of the solution is now given by
\begin{equation}
Q = 4\pi \kappa m ,    ~~m=1,2,3...
\end{equation}
The nonabelian magnetic field of the solution can be expressed as
\begin{equation}
F_{+-} = m( |m-1\rangle\langle{m-1}|-|m\rangle\langle{m}|) \textrm{diag}(0,...,1),
\end{equation}
and thus the net magnetic flux vanishes.


To summarize, we consider solitonic solutions in the noncommutative $U(N)$ Chern-Simons gauge theory coupled
to a  scalar field in the fundamental and adjoint representations.
When coupled to fundamental scalar, we obtain solutions by embedding the  solutions
of noncommutative abelian Chern-Simons-Higgs model.
With adjoint scalar, we get a class of particular solutions  whose commutative counterparts do not exist.
Full analysis of   general BPS solutions with adjoint scalar needs further investigation.

\section*{Acknowledgments}
This work was supported by
the Korea Science and Engineering Foundation(KOSEF) grant
funded by the Korea government(MEST), R01-2008-000-21026-0(E. C.-Y. and D. L.),
and by the National Research Foundation(NRF) of Korea grant
funded by the Korea government(MEST), 2009-0075129 (E. C.-Y. and K. K.),
and by the Korea Research Foundation grant funded by
the Korea government(MOEHRD), KRF-2008-314-C00063 (Y. L.).


\end{document}